# Deep Medical Image Analysis with Representation Learning and Neuromorphic Computing

Getty, N.[1,2], Brettin, T.[1,3], Jin, D.[2], Stevens. R.[1,3], Xia, F.[1,3]

[1]Argonne National Laboratory, Lemont, IL, U.S.A.

[2]Illinois Institute of Technology, Chicago, IL, U.S.A.

[3]University of Chicago, Chicago, IL, U.S.A.

## 1. Overview

Deep learning is increasingly used in medical imaging, improving many steps of the processing chain, from acquisition to segmentation and anomaly detection to outcome prediction. Yet significant challenges remain: (1) image-based diagnosis depends on the spatial relationships between local patterns, something convolution and pooling often do not capture adequately; (2) data augmentation, the de facto method for learning 3D pose invariance, requires exponentially many points to achieve robust improvement; (3) labeled medical images are much less abundant than unlabeled ones, especially for heterogenous pathological cases; and (4) scanning technologies such as magnetic resonance imaging (MRI) can be slow and costly, generally without online learning abilities to focus on regions of clinical interest. To address these challenges, novel algorithmic and hardware approaches are needed for deep learning to reach its full potential in medical imaging.

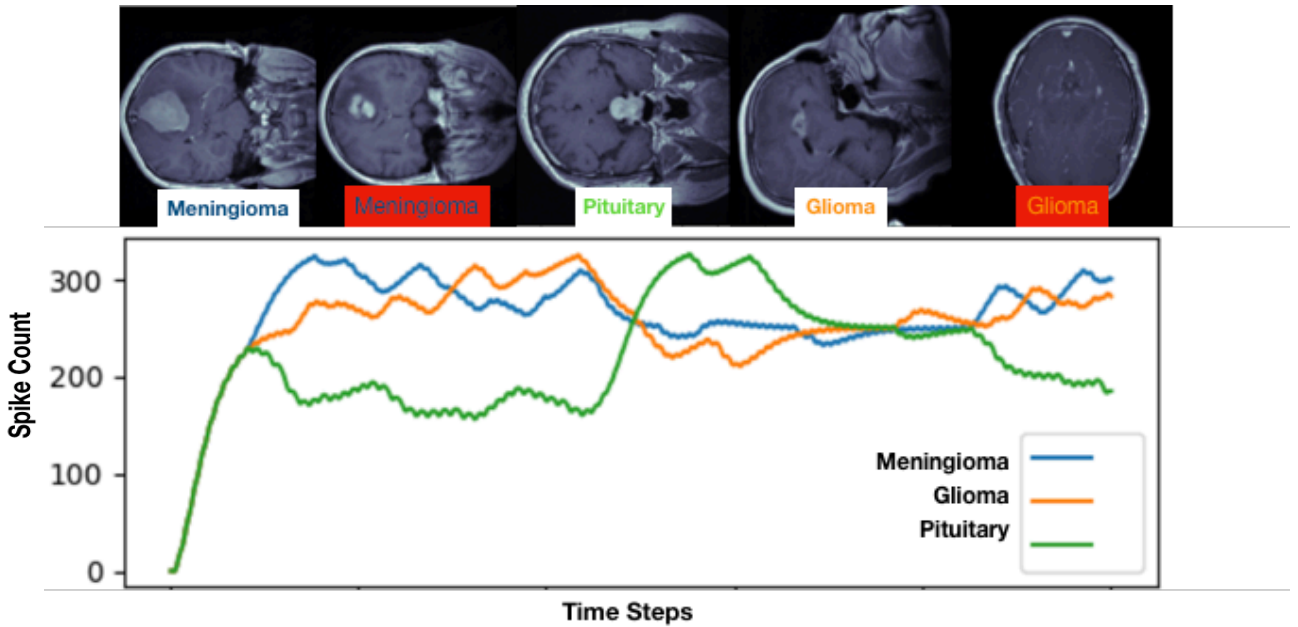

***Figure 1*** *Brain tumor image classification with a Spiking Neural Network on Intel's neuromorphic hardware. The spike count of the final output neurons is used for prediction, misclassifications highlighted in red.*

We explore three representative lines of research and demonstrate the utility of our methods on a classification benchmark of brain cancer MRI data. First, we present a capsule network that explicitly learns a representation robust to rotation and affine transformation. This model requires less training data and outperforms both the original convolutional baseline and a previous capsule network implementation. Second, we leverage the latest domain adaptation techniques to achieve a new state-of-the-art accuracy. Our experiments show that non-medical images can be used to improve model performance. Finally, we design a spiking neural network trained on the Intel Loihi neuromorphic chip (Fig. 1 shows an inference snapshot). This model consumes much lower power while achieving reasonable accuracy given model reduction. We posit that more research in this direction combining hardware and learning advancements will power future medical imaging (on-device AI, few-shot prediction, adaptive scanning).

## 2. Results

We tested our methods on a representative medical imaging dataset [1]. This benchmark contains 3,064 MRI slices of 233 patients diagnosed with one of the three brain tumor types: meningioma (23%), glioma (47%), and pituitary tumor (30%). We compare our results with a convolutional network baseline and a previous state-of-the-art capsule network implementation.

*Table 1* Comparing our three approaches with two previous models

| Classification Model | Test Accuracy | Sample Efficiency (Test accuracy with 10% training data) | Energy Efficiency (Joules/Inference) |
|---|---|---|---|
| ResNet trained from scratch | 83.70% | 61.90% | 0.1190 |
| Previous CapsNet [1] | 85.60% | N/A | N/A |
| Our CapsNet | 89.30% | **75.30%** | 0.1745 |
| Pretrained ResNet | **92.40%** | 64.70% | 0.1190 |
| SpikingNet on Intel Loihi chip | 85.6% | N/A | 0.0016-0.0052* |

Table 1 summarizes model performance differences in terms of test accuracy (fraction of accurately classified samples), sample efficiency, and energy efficiency. We stratified the images both by patient and for a balanced tumor classification and used a conservative 30% split for the test set. Our domain adaptation approach, implemented with a pretrained residual network, achieved a new state-of-the-art accuracy of 92.4%. Among models trained from scratch, our capsule network had the highest classification accuracy. To assess sample efficiency, we measured model performance trained on a fraction of the data ranging from 10% to 50%. In these low data scenarios, the capsule network consistently outperformed other models by a large margin (over 10% with 10% training data).

*Table 2* Inference speed of each approach

|  | Capsule | ResNet | SNN |
|---|---|---|---|
| Inferences/Second | 143 | **324** | 106 |

To calculate energy consumed per inference, we divided the average dynamic energy usage by the number of inferences per second. The minimum Loihi estimate is calculated with the finding from Blouw et. al. [10] that a network run on Loihi was 109x more energy efficient than GPU. The maximal estimate is derived by calculating the dynamic energy usage per neurocore and again dividing by the inferences per second. While Blouw et. al. fit their model on 8 cores, we required 55 cores and inference of our network was 3x slower than their keyword spotting application. This may be considered an upper bound, as it does not account for the fact that when a neurocore is inactive it does not consume any power. Speed (Table 2) and energy efficiency of the spiking approach could be increased by reducing the number of timesteps used for each classification. Halving the number of timesteps results in a 5.7% accuracy trade off.

## 3. Methods and Discussion

### Spatial representation learning with Capsule Networks

Capsule networks [2] are a divergence from typical convolutional neural networks which rely on pooling to move from simple features detected early in the network to higher level features. What we lose in pooling is spatial resolution. While other techniques such as dilated convolution attempt to assuage this loss, capsules focus on explicitly capturing spatial hierarchies between simple and complex patterns through dynamic routing. The advantage is some level of representational invariance to image transformations including pose, lighting or deformation. This advantage makes capsules a good fit for medical images with diffuse, diverse abnormalities as well as variances from both measuring devices and individual differences.

*Table 3* *Cross validation experiment for capsule networks*

| Metric | Fold 1 | Fold 2 | Fold 3 | Fold 4 | Fold 5 | Average |
|---|---|---|---|---|---|---|
| **Accuracy** | 0.886 | 0.910 | 0.852 | 0.872 | 0.893 | 0.883 |
| **MCC** | 0.822 | 0.857 | 0.776 | 0.793 | 0.826 | 0.815 |
| **F1 Score** | 0.890 | 0.910 | 0.850 | 0.870 | 0.900 | 0.884 |

*Matthews correlation coefficient is a summary metric for the confusion matrix and generalizes to unbalanced multi-class problems.*

Capsule networks [2] are distinct in three key ways: (1) layers are made up of many groups of neurons called *capsules* which output a vector whose length represents probability and whose orientation encodes spatial, pose, or other instantiation parameters.; (2) a dynamic, hierarchical parse tree is generated by routing the outputs of lower-layer capsules to higher, abstract capsules which agree with their output. With this routing-by-agreement, parts of a whole may be cohesively organized; (3) the activation vectors of the final output capsules are fed to a decoder network which attempts to reconstruct the original image. This acts to regularize the network and encourage the encoding of spatial, pose, and transformation parameters. This reconstruction loss is directly involved during training of the network.

*Table 4 Tumor specific classification performance*

| Tumor Type | Precision | Recall | F1 |
|---|---|---|---|
| Meningioma | 0.736 | 0.778 | 0.746 |
| Glioma | 0.912 | 0.874 | 0.892 |
| Pituitary | 0.914 | 0.942 | 0.938 |

We replicated the capsule network architecture [1] by Guo [3] and optimized the preprocessing and model parameters for the raw MRI tumor images rather than the derived tumor segmentation masks. The architecture matches that in [1] with an input convolutional layer with 256 filters and kernel size of 9, an additional dropout layer, a 32 channel PrimaryCapsule layer with 8 dimensional capsules, kernel size of 9 and stride of 2, another additional dropout layer, and 3 output capsules (number of classes) with dimension 16 and 3 dynamic routing iterations. Furthermore, we utilized learning rate decay, larger batch sizes and longer training time.

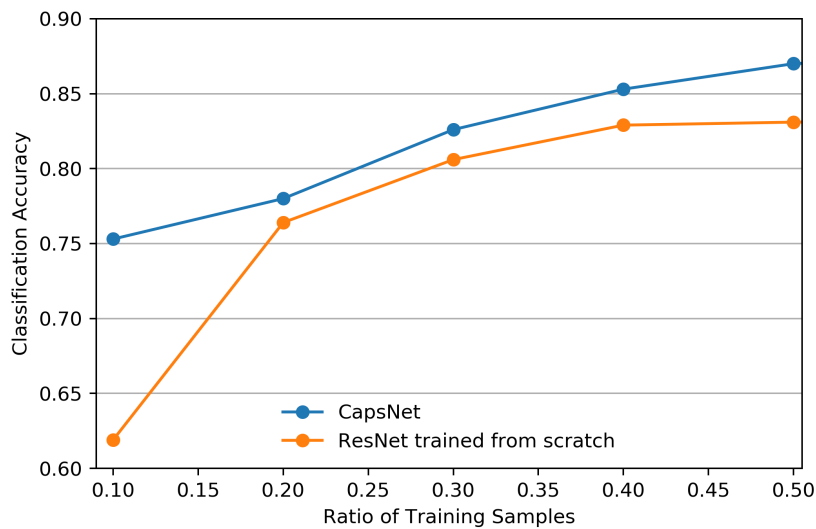

*Figure 2 Low data learning: classification accuracy as a function of the fraction of training examples*

Ultimately our trained model marks a significant performance improvement on the raw MRI images, requiring no manually segmentations of the tumor region of interest. The model clearly generalizes and is robust to different partitions of the dataset split by patient and stratified by tumor type (Table 3). Unsurprisingly, the tumor type with the lowest available number of samples had modestly worse performance (Table 4 and Figure 3). Notably, meningioma tumors may occur anywhere between the skull and surface of the brain and may appear to be within the brain depending on the viewing plane and slice location on the image. The second image in Figure 1 is a good example of this, where the model incorrectly classifies meningioma as glioma for a case that in 2D does appear to be inside the brain rather than on the surface. While the accuracy did not surpass the pretrained ResNet model, it outperformed the ResNet model trained from scratch. The capsule model is also data efficient, with performance degrading less substantially as the amount of labeled data decreases (Fig. 2).

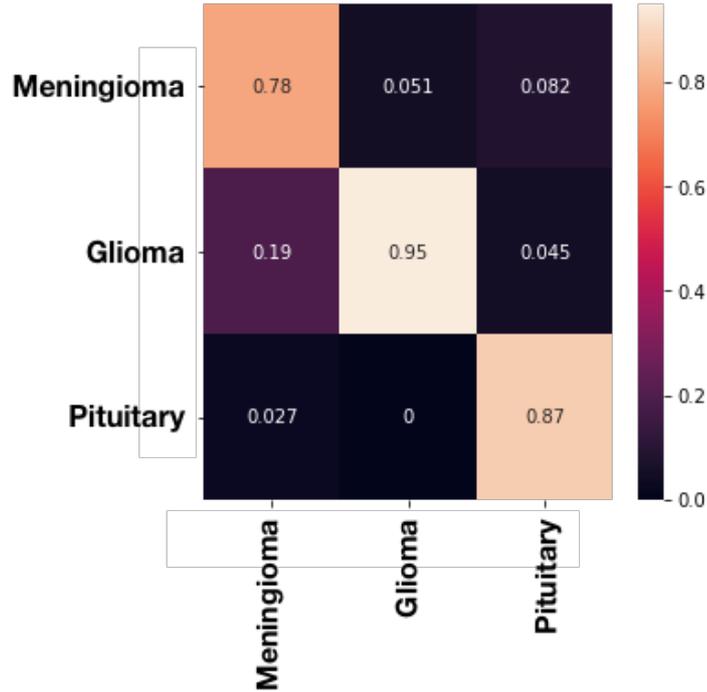

*Figure 3* Confusion matrix for the predicted tumor types.

Another advantage of capsule networks is explainability. By manipulating the vector output of the capsule layer and reconstructing the image with the jointly learned decoder, one can visualize what each capsule is learning. This is especially vital in clinical settings where transparency may improve both doctor and patient trust and has the potential to uncover previously unremarkable image-based biomarkers that are relevant for diagnosis and prognosis. By analyzing the confusion matrix (Fig. 3), we may reconstruct images from classes that are more often misclassified to understand in what way the model is failing.

**Transfer learning**

Our best model in terms of classification accuracy is a 50-layer convolutional residual network [16], improving on the previous state-of-the-art by 6.8%. The training process implemented using the FastAI [14] framework, is as follows: we started with the pretrained weights for ImageNet, and first trained only the top fully-connected layer while keeping the residual blocks frozen. At this stage, the model reached 87.1% test accuracy. We then unfroze all layers and assigned them differential starting learning rates, giving more freedom to the higher layers. A second round of training with cyclical learning rates [15] boosted the final test accuracy to 92.4%.

The key component to residual networks is the introduction of a skip-connection from a previous layer's output to the next layer. Empirically, these connections enable the design of much deeper neural networks which avoid degradation due to accuracy saturation [16]. By fitting a residual mapping using skip-connections, any degradation is avoided, and deeper networks may indeed offer improved performance. In pretraining the network and fine-tuning on a new domain, our model may in effect learn the basics of image classification which should be common for any dataset. The most powerful capability of neural networks is the automatic detection and efficient representation of increasingly higher-order features.

In an ablation experiment to quantify the value of pretrained weights, we trained the same model architecture from scratch. This baseline model achieved 83.7% test accuracy. This comparison points to an 8% boost from domain adaptation. Although ImageNet contains no medical images, the learned lower-level features were able to generalize to the brain scans. Recent efforts in building large-scale medical imaging databases will likely provide an even better base for transfer learning.

**Spiking network on neuromorphic device**

Despite the tremendous success of AI, we understand very little about the human intelligence algorithm. Unlike the point neuron model used in deep learning, real neurons are connected to thousands of excitatory synapses hypothesized to recognize multiple independent patterns. Neuromorphic computing aims to get one step closer to how brain works with spiking signals and local learning rules. Rather than employing a non-linear activation function, spiking neurons must reach an activation potential before generating an output spike to forward connections and resetting. Learning rules are assigned to synapses, defining how the weights are computed during learning as a function of the pre and post synaptic traces. Neuromorphic hardware is an asynchronous design architecture specifically for implementing spiking based neural networks.

Non-Von Neumann hardware such as neuromorphic hardware is a clear candidate for enabling efficient learning for edge computation. Motivation for such hardware includes sensor networks as well as for acceleration of machine learning algorithms. Similarly, spiking neural networks are emerging as a potential next generation of artificial neural networks. Most ongoing research is focused on developing and adapting more brain inspired algorithms to run on neuro hardware. Many of the current applications of the hardware are aptly in the robotics domain. However, the energy efficiency, processing speed and scalability of such systems make this hardware a candidate for many applications, especially those which require edge processing.

What currently sets Loihi apart from other neuromorphic platforms is the online learning capacity and system scalability. Compared with another well-known neuromorphic system, IBM's TrueNorth [13], has 4096 cores each consisting of a 256x256 crossbar of neurons/synaptic connections. As opposed to Loihi's pulse processors, TrueNorth employs a time-multiplexed processor in each core. Where Loihi provides 3x higher synaptic density, TrueNorth provides higher neuron density. Synaptic delays and 8-bit precision are supported. TrueNorth has implemented a 16-chip scaled design but has not provided an upper limit for potential scalability. One limit however is a neuron may only communicate up to 4 chips away in either direction in the mesh.

We implemented four spiking networks on and off chip using Nengo [5] the Loihi SDK and SNN Toolbox [11]: (1) a single layer image classifier, trained on chip, which encodes every pixel's intensity via a random Poisson spike generation process; (2) a converted convolutional neural network which successfully ran on the chip simulator but resulted in the actual chip timing up; (3) a spiking fully connected network which dramatically reduces the total number of neurons and processing time by first preprocessing images using dimensionality decomposition; (4) a continuous valued convolutional neural network converted to a SNN (Table 1) using SNN Toolbox. The loss in accuracy observed for the converted CNN is largely a result of reducing the input size and complexity of the model. A loosely adopted MobileNet [12] CNN architecture with fewer convolutional filters achieved 86.36%, meaning the actual error introduced by conversion was 0.76%. Currently, only 2D convolutional, pooling, dense, flatten and resize/padding type layers are supported, precluding direct conversion of ResNet or Capsule Network architectures.

## 4. Conclusions

We have demonstrated three different approaches to improve medical imaging analysis tackling data efficiency, transfer learning and on-device learning. We posit the application of neuromorphic computing coupled with new learning algorithms will stimulate advancements in imaging technologies. Today's MRI scans are slow and often require patients to be still for extended periods. Acceleration and enhancement may be realized at several steps of the process: (1) real-time adaptive scanning implemented to increase spatial and temporal resolution at automatically detected regions of interest; (2) accelerated data acquisition by intelligent compressive sampling and reconstructive techniques to reduce number of slices and required TR/TE times per slice; (3) faster algorithm implementation of the final image reconstruction; and (4) real-time, few-shot learning for image enhancement techniques including motion correction, anomaly detection and anatomical segmentation. The benchmarking and exploration of novel software and hardware here may serve as a framework for countless other applications with similar considerations and challenges.